\documentclass[]{iopart}

\usepackage{amsmath1,amssymb,amscd,amsthm}
\usepackage{epsfig}

\newtheorem{theorem}{Theorem}
\newtheorem{lemma}{Lemma}
\newtheorem{proposition}{Proposition}

\newtheorem{corollary}{Corollary}

\newtheorem{claim}{Claim}

\newcommand{\norm}[2]{{\left\| #1 \right\|}_{#2}}
\newcommand{\f}[2]{\frac{#1}{#2}}
\newcommand{\dpr}[2]{\langle #1,#2 \rangle}

\newcommand{\ga}{\gamma}

\newcommand{\ve}{\varepsilon}

\newcommand{\rd}{{\mathbb R}^d}

\newcommand{\intl}{\int\limits}

\newcommand{\p}{\partial}

\newcommand{\beq}{\begin{equation}}
\newcommand{\eeq}{\end{equation}}
\newcommand{\beqna}{\begin{eqnarray*}}
\newcommand{\eeqna}{\end{eqnarray*}}
\newcommand{\beqn}{\begin{equation*}}
\newcommand{\eeqn}{\end{equation*}}
\newcommand{\bp}{\begin{proof}}
\newcommand{\ep}{\end{proof}}
\newcommand{\bprop}{\begin{proposition}}
\newcommand{\eprop}{\end{proposition}}
\newcommand{\bt}{\begin{theorem}}
\newcommand{\et}{\end{theorem}}
\newcommand{\bex}{\begin{Example}}
\newcommand{\eex}{\end{Example}}
\newcommand{\bc}{\begin{corollary}}
\newcommand{\ec}{\end{corollary}}
\newcommand{\bcl}{\begin{claim}}
\newcommand{\ecl}{\end{claim}}
\newcommand{\bl}{\begin{lemma}}
\newcommand{\el}{\end{lemma}}

\begin{document}

\title[Nonlinearity Management in Higher Dimensions]
{Nonlinearity Management in Higher Dimensions}

\author{Panayotis G. Kevrekidis}

\address{
Department of Mathematics and  Statistics,
University of Massachusetts,
Amherst, MA 01003}

\author{Atanas Stefanov}

\address{
Department of Mathematics, 
University of Kansas,
1460 Jayhawk Blvd, Lawrence, KS 66045--7523}

\author{Dmitry E. Pelinovsky}

\address{
Department of Mathematics,
McMaster University,
Hamilton, Ontario,
Canada, L8S 4K1}



\begin{abstract}
In the present short communication, we revisit nonlinearity management of the
time-periodic nonlinear Schr\"{o}dinger equation and the related
averaging procedure. We prove that the averaged nonlinear
Schr{\"o}dinger equation does not support the blow-up of solutions
in higher dimensions, independently of the strength in the
nonlinearity coefficient variance. This conclusion agrees with
earlier works in the case of strong nonlinearity management but
contradicts those in the case of weak nonlinearity management.
The apparent discrepancy is explained by the divergence of the averaging
procedure in the limit of weak nonlinearity management.
\end{abstract}

\maketitle
\date{today}

\section{Introduction}

In the past few years, there has been a large volume of literature
regarding the applications of the nonlinear Schr{\"o}dinger
equation (NLS) in the presence of the so-called {\it nonlinearity
management} (often referred to also as Feshbach resonance
management). The NLS is a prototypical dispersive nonlinear wave
equation of the form:
\begin{eqnarray}
i u_t = - \Delta u + \Gamma(t) |u|^2 u + V(x) u, \label{eqn1}
\end{eqnarray}
where $u(x,t)$ is a complex envelope field, $V(x) \geq 0$ is an
external potential, $\Gamma(t)$ is a time-periodic nonlinearity
coefficient, and $\Delta$ is the Laplacian operator with $x \in
\mathbb{R}^d$, $d \geq 1$.

Nonlinearity management arises in applications in optics for
transverse beam propagation in layered optical media \cite{mal1}, as
well as in atomic physics for the Feshbach resonance of the
scattering length of inter-atomic interactions in Bose-Einstein
condensates (BECs) \cite{inouye}. In the latter case, the periodic
variation of $\Gamma(t)$ through an external magnetic field has been
used as a means of producing robust matter-wave breathers in
quasi-one-dimensional BECs \cite{KTFM}. It has also been suggested
that the nonlinearity management may prevent collapse-type phenomena
in higher dimensions \cite{abdul}. Theoretical studies of the
nonlinearity management were performed with a reduction of the
time-periodic PDE problem (\ref{eqn1}) to a time-periodic ODE
problem using a variational method \cite{saito} and a moment method
\cite{perez}.

The physical relevance of the time-periodic NLS equation, as
evidenced by the above works, led to further developments in
analysis of the PDE problem (\ref{eqn1}). As an example, the success
of the averaging theory \cite{GS} for optical solitons in the
presence of {\it strong dispersion management}, led to an analogous
development for {\it strong nonlinearity management} produced
originally in \cite{PKF1} and systematized in \cite{PZ1}. The
time-periodic NLS equation (\ref{eqn1}) is averaged in the limit
$\epsilon \to 0$, where $\epsilon$ measures a short period and the large
variation of $\Gamma(t)$ in the decomposition
\begin{equation}
\Gamma = \gamma_0 + \frac{1}{\epsilon} \gamma\left(
\frac{t}{\epsilon}\right), \label{eqn2aa}
\end{equation}
where $\gamma(\tau)$, $\tau = t/\epsilon$ has a unit period and zero
mean. After the averaging procedure, the time-periodic PDE problem
(\ref{eqn1}) is governed by the averaged NLS equation,
\begin{equation}
i w_t = - \Delta w + \gamma_0 |w|^2 w + V(x) w - \sigma^2 \left(
\left| \nabla |w|^2 \right|^2 + 2 |w|^2 \Delta |w|^2 \right) w,
\label{eqn0}
\end{equation}
where
\begin{equation}
\sigma^2 = \int_0^1 \gamma_{-1}^2(\tau) d \tau,
\end{equation}
and $\gamma_{-1}(\tau)$ is the mean-zero anti-derivative of
$\gamma(\tau)$. Derivation and local well-posedness of solutions of
the averaged NLS equation (\ref{eqn0}) in function space
$H^{\infty}(\mathbb{R})$ are reviewed for $d = 1$ in \cite{PZ1}.

In the present paper, we consider global well-posedness of solutions
of the averaged NLS equation (\ref{eqn0}) in the energy space
$H^1(\mathbb{R}^d)$ for $d \geq 1$. The use of $H^1$ and $d \geq 1$
seems more appropriate for physical applications of the averaged
model (\ref{eqn0}). In particular, we address the question whether
the averaged NLS equation with a nonlinearity management $(\sigma^2
> 0)$ arrests the blowup of solutions of the NLS equation in two and
three dimensions that would occur if no nonlinearity management was
applied $(\sigma^2 = 0$). We show that the averaged NLS equation is
globally well-posed and no blowup of solutions occurs for $\sigma^2
> 0$. This is demonstrated initially, in Section 2, from the point
of view of scaling arguments. The statement is subsequently proved
using rigorous estimates in Section 3. In Section 4 we compare the
above conclusion and earlier works where possibilities of blowup of
solutions of the full time-periodic NLS equation (\ref{eqn1}) have
been reported. Section 5 summarizes our findings.

\section{Formal Scaling Arguments}

The averaged NLS equation (\ref{eqn0}) has a standard Hamiltonian
form (see \cite{PZ1}) with the Hamiltonian functional:
\begin{eqnarray}
H(w) = \intl_{\rd} \left( |\nabla w|^2 + \frac{\gamma_0}{2} |w|^4 +
V(x) |w|^2 + \sigma^2 |w|^2 \left|\nabla |w|^2 \right|^2 \right) dx.
\label{eqn2}
\end{eqnarray}
Due to the gauge invariance, the averaged NLS equation (\ref{eqn0})
also conserves the squared $L^2$ norm:
\begin{equation}
\label{eqn2a} P(w) = \intl_{\rd} |w|^2 dx.
\end{equation}
Solitary wave solutions of the averaged NLS equation (\ref{eqn0})
are critical points of $H(w)$ at the level set of fixed values of
$P(w)$.

Using formal scaling arguments \cite{derrick} (see also the
review of Ref. 
\cite{ZRK}), we consider a two-parameter family of dilatations:
\begin{equation}
\label{eqn4a} w = b W(a x),
\end{equation}
where $(a,b)$ are parameters and $W(\xi)$ is a suitable function of
$\xi = a x$. The squared $L^2$ norm (\ref{eqn2a}) is preserved by
the dilatations (\ref{eqn4a}) whenever $b=a^{d/2}$. The Hamiltonian
(\ref{eqn2}) at the dilatations (\ref{eqn4a}) is scaled as a
function of parameter $a > 0$:
\begin{eqnarray}
H(a) = I_0(a) + a^2 I_1 + \gamma_0 a^d I_2 + \sigma^2 a^{2 d+2}
I_3, \label{eqn4}
\end{eqnarray}
where
\begin{eqnarray*}
I_1 = \intl_{\rd} |\nabla W|^2 d \xi, \quad I_2 = \frac{1}{2}
\intl_{\rd} |W|^4 d \xi, \quad I_3 = \intl_{\rd} |W|^2 \left|\nabla
|W|^2 \right|^2 d \xi.
\end{eqnarray*}
and
$$
I_0(a) = \intl_{\rd} V\left(\frac{\xi}{a}\right) |W|^2 d\xi.
$$
Let us consider the case of no nonlinearity management and no
external potential, when $\sigma^2 = 0$ and $V(x) = 0$. It follows
from (\ref{eqn4}) that the Hamiltonian function $H(a)$ is positive
definite in the defocusing case, when $\gamma_0 > 0$. In the
focusing case, when $\gamma_0 < 0$, the Hamiltonian function $H(a)$
is bounded from below for $d = 1$ and $d = 2$, $\gamma_{\rm cr} <
\gamma_0 < 0$ and is unbounded from below for $d = 2$, $\gamma_0 <
\gamma_{\rm cr}$ and $d = 3$, where
$$
\gamma_{\rm cr} = - \frac{I_1}{I_2}.
$$
When $H(a)$ is unbounded from below as $a \to \infty$, the critical
points of $H(w)$ at a fixed value of $P(w)$ (i.e., solitary wave
solutions) can not be stable for small width $a^{-1}$ and
instability of solitary waves implies a blowup of localized initial
data in the time evolution of the cubic NLS equation (see \cite{ZRK}
for details).

When the nonlinearity management is applied, the last term in the
decomposition (\ref{eqn4}) always dominates and it preserves the
boundness of $H(a)$ from below for any $\sigma^2 > 0$. This
indicates on the level of formal scaling arguments that the blowup
of solutions is arrested by the nonlinearity management term in the
averaged NLS equation (\ref{eqn0}). We shall prove this conjecture
with rigorous analysis of well-posedness of solutions. We also note
that the first term in the decomposition (\ref{eqn4}) does not
change the conclusions above if $V(x)$ is a smooth non-negative
potential, such that $I_0(a) \geq 0$. Typical examples of $V(x)$ are
parabolic magnetic traps, when $V \sim x^2$, and periodic optical
lattices, when $V \sim \sin^2(k_0 x)$.

\section{Rigorous Analysis of Well-Posedness}

The rigorous analysis of the local well-posedness of the averaged
NLS equation \eqref{eqn0} is not  a trivial task. In fact, to the
best of our knowledge, one cannot verify even the local existence
and uniqueness of solutions to this problem. The problem has been
considered in one dimension $d = 1$ by Poppenberg where a local
well-posedness result with data in $H^\infty$ was established
\cite{poppenberg}. In higher dimensions $d \geq 2$, one needs to
require that the initial data be in $H^{s,m}$ for sufficiently large
$s$,$m$, that is $\int_{\mathbf R^n} (1+|x|)^{2m} (|w_0|^2+|\p^s
w_0|)^2 dx<\infty$ (see \cite{kenig}). In addition, one needs to
assume a ``non-trapping'' condition on the symbol of the second
order operator. This is a geometric condition, which depends on the
profile of the initial data (see \cite{kenig} for details).

To summarize, we cannot state a precise condition under which the
averaged NLS equation \eqref{eqn0} has a (local) solution that
preserves values of $P$ and $H$ constant in time. We will however
show, that whenever such a local solution exists for a short time $0
< t < t_0 < \infty$, it can be extended globally for all $t > 0$.

To that end, we represent the Hamiltonian $H(w)$ in the form:
$$
H (w) = H_1(w) + \gamma_0 H_2(w),
$$
where
$$
H_1(w)= \intl_{\rd} \left(|\nabla w|^2+ V(x) |w|^2 + \sigma^2 |w|^2
|\nabla|w|^2|^2\right) dx \geq 0
$$
and
$$
H_2(w) =  \frac{1}{2} \intl_{\rd} |w|^4 dx \geq 0.
$$
We consider the focusing case $\gamma_0 < 0$ and prove that $H_1(w)$
and $H_2(w)$ are bounded by the two conserved quantities $H(w)$ and
$P(w)$.

First, we quote a variant  of the Gagliardo-Nirenberg inequality
(see (1.1.16) on p. 15 in \cite{Grafakos}).

\begin{lemma}
For all $1\leq p,q,r\leq \infty$, $\theta\in (0,1)$ and
$r^{-1}=\theta p^{-1}+(1-\theta) q^{-1}$, it is true for every
function $f(x)$ on $x \in \mathbb{R}^d$ that
\begin{equation}
\label{GN} \norm{f}{L^r} \leq \norm{f}{L^p}^\theta
\norm{f}{L^q}^{1-\theta}.
\end{equation}
\end{lemma}
\noindent

Next, we define and use the Fourier transform and its inverse for a
function $f(x)$ on $x \in \mathbb{R}^d$:
\begin{eqnarray}
\hat{f}(\xi)=\intl_{\rd} f(x) e^{-2\pi i \dpr{x}{\xi}} dx, \quad
f(x)=\intl_{\rd} \hat{f}(\xi) e^{2\pi i \dpr{x}{\xi}} d\xi.
\end{eqnarray}
The Plancherel's formula gives $\norm{f}{L^2}=\|\hat{f}\|_{L^2}$
 and  the inequality
$\norm{f}{L^\infty}\leq \|\hat{f}\|_{L^1}$ is immediate from the definition.

Let $\chi(s)$ be a $C_0^\infty(\mathbb{R}_+)$ function with a
compact support on $s \in [0,2]$, such that $\chi(s) = 1$ on $s \in
[0,1]$. Let $\chi_d(x)$ be a $C_0^\infty(\mathbb{R}^d)$ function,
such that $\chi_d(x) = \chi(|x|)$. The (smooth) Fourier multiplier
$P_{<N}$ is defined for every positive $N$ by
$\widehat{P_{<N}f}(\xi)=\chi_d(\xi/N)\hat{f}(\xi)$. Equivalently,
\begin{equation}
P_{<N}f(x)=N^d \intl_{\rd} \widehat{\chi_d}(N(x-y))f(y)
dy.
\end{equation}
Let $P_{>N} = Id-P_{<N}$. Then both $P_{>N}, P_{<N}$ are
self-adjoint, bounded on $L^2$ operators. The following statement is
a modification of the Sobolev embedding theorem.

\begin{lemma}
\label{le:1} There exists a constant $C_d > 0$, which depends only
on the dimension $d \geq 1$, so that it is true for every function
$f(x)$ on $x \in \mathbb{R}^d$ that
\begin{equation}
\label{eq:1} \norm{f}{L^2} \leq C_d \left( \norm{\nabla f}{L^2}+
\norm{f}{L^1} \right).
\end{equation}
\end{lemma}
\begin{proof}
Since $f = P_{<1}f + P_{>1}f$, it will suffice to show that
\begin{eqnarray}
\label{eq:7}
& &\norm{P_{<1} f}{L^2}\leq C_d \norm{f}{L^1} \\
\label{eq:8} & & \norm{P_{>1} f}{L^2}\leq C \norm{\nabla f}{L^2}
\end{eqnarray}
The bound \eqref{eq:8} follows simply by the Plancherel's theorem
and $\widehat{\p_j f}(\xi)=2\pi i \xi_j\hat{f}(\xi)$. Indeed,
\begin{eqnarray*}
\norm{P_{>1} f}{L^2}^2 &\leq& \intl_{|\xi|>1} |\hat{f}(\xi)|^2
d\xi\leq \intl_{|\xi|>1} |\xi|^2|\hat{f}(\xi)|^2d\xi=\f{1}{4\pi^2}
\intl_{|\xi|>1} |\widehat{\nabla f}(\xi)|^2 d\xi\leq \\
& &\leq \f{1}{4\pi^2} \intl_{\rd} |\widehat{\nabla f}(\xi)|^2 d\xi
= \f{1}{4\pi^2} \norm{\nabla f}{L^2}^2.
\end{eqnarray*}
By duality, the bound \eqref{eq:7} follows from the estimate
$\norm{P_{<1} f}{L^\infty}\leq C_d \norm{f}{L^2}$. That is trivial
as well, since
$$
\norm{P_{<1} f}{L^\infty}\leq \norm{\widehat{P_{<1} f}}{L^1}\leq
\intl_{|\xi|\leq 2} |\hat{f}(\xi)|d\xi\leq C_d
 (\intl_{|\xi|\leq 2} |\hat{f}(\xi)|^2 d\xi)^{1/2}\leq  C_d
\norm{f}{L^2}.
$$
The positive constant $C_d$ can be taken to be the square root of
the volume of the ball in $\mathbb{R}^d$ with radius $2$.
\end{proof}

The central result of our analysis is the following theorem.

\begin{theorem}
\label{theo:1} There exist $\ve(d)>0$ and $C(\ve,d,\sigma) > 0$,
so that for every $0<\ve<\ve(d)$, it is true for every function
$\phi(x)$ on $x \in \mathbb{R}^d$ that
\begin{equation}
\label{eq:2} \norm{\phi}{L^4}^4 \leq \ve H_1(\phi) +
C(\ve,d,\sigma) \left(\norm{\phi}{L^2}^{2}+\norm{\phi}{L^2}^{4}
\right).
\end{equation}
\end{theorem}

\begin{proof}
Let $f = h^{3/2}$ with $h(x) > 0$ in \eqref{eq:1} and obtain
$$
\intl_{\rd} h^3 dx \leq C_d^2 \left( \frac{3}{2} \left( \intl_{\rd}
h |\nabla h|^2 dx \right)^{1/2} + \intl_{\rd} h^{3/2} dx \right)^2.
$$
Next, we set $h=|\phi|^2$ and obtain
\begin{eqnarray*}
& &
\intl_{\rd} |\phi|^6 dx\leq C_d^2 \left( \frac{3}{2} \left(
\intl_{\rd} |\phi|^2 |\nabla |\phi|^2|^2 dx \right)^{1/2} +
\intl_{\rd} |\phi|^{3} dx \right)^2 \\
& & \leq C_d^2\left(\frac{9}{2} \intl_{\rd} |\phi|^2 |\nabla |\phi|^2|^2 dx+ 2
\left(\intl_{\rd} |\phi|^{3} dx\right)^2\right) \\
& &\leq  C_{d,\sigma} (H_1(\phi)+ \norm{\phi}{L^3}^6),
\end{eqnarray*}
for some positive constant $C_{d,\sigma}$. We have used here that
$$
\left( \sqrt{a} + b \right)^2 \leq 2\left( a+ b^2 \right).
$$
By the Gagliardo-Nirenberg inequality (\ref{GN}), we have
$\norm{\phi}{L^3}\leq \norm{\phi}{L^2}^{1/2}\norm{\phi}{L^6}^{1/2}$,
such that the last inequality is rewritten in the form:
\begin{equation}
\label{eq:3} \norm{\phi}{L^6}^6\leq C_{d,\sigma} H_1(\phi)+ C_d^2
\norm{\phi}{L^6}^3 \norm{\phi}{L^2}^3\leq C_{d,\sigma} H_1(\phi) +
C_d^2 \left( \ve\norm{\phi}{L^6}^6+ \f{1}{4\ve} \norm{\phi}{L^2}^6
\right),
\end{equation}
where in the last line we have used the Cauchy-Schwartz inequality:
\begin{equation}
\label{CS} \forall \ve > 0 : \;\; ab \leq \ve a^2 + \frac{b^2}{4
\ve}.
\end{equation}
Let $\ve < \ve(d)$, where $2 C_d^2 \ve(d) = 1$. Then, the term
$\norm{\phi}{L^6}^6$ can be estimated from the bound \eqref{eq:3} as
follows:
\begin{equation}
\label{eq:5} \norm{\phi}{L^6}^6 \leq \tilde{C}_{d,\sigma} H_1(\phi)+
\tilde{C}_d \norm{\phi}{L^2}^6
\end{equation}
for some constants $\tilde{C}_{d,\sigma} > 0$ and $\tilde{C}_d
> 0$. By the Gagliardo-Nirenberg inequality (\ref{GN}), we have $\norm{\phi}{L^4}\leq
\norm{\phi}{L^6}^{3/4}\norm{\phi}{L^2}^{1/4}$, such that the upper
bound for $\norm{\phi}{L^4}$ follows from (\ref{CS}) and
\eqref{eq:5}:
$$
\norm{\phi}{L^4}^4 \leq \norm{\phi}{L^2}(\hat{C}_{d,\sigma}
\sqrt{H_1(\phi)}+ \hat{C}_d \norm{\phi}{L^2}^3) \leq
\frac{\hat{C}_{d,\sigma}}{4 \ve} \norm{\phi}{L^2}^2 + \hat{C}_d
\norm{\phi}{L^2}^4 + \ve H_1(\phi),
$$
which is the desired upper bound \eqref{eq:2}.
\end{proof}

As a corollary of the main theorem, we pick $\ve=\ve(d)/2$ and
immediately obtain the following upper bounds.

\begin{corollary} There exists constants
$C_1 > 0$ and $C_2 > 0$ that depend on $d$,$\ga_0$,$\sigma$, so that
\begin{equation}
\label{eq:6} H_1(w)\leq C_1 \left( H(w) + P(w) + P^2(w) \right)
\end{equation}
and
\begin{equation}
H_2(w) \leq C_2 \left( H(w) + P(w) + P^2(w) \right).
\end{equation}
\end{corollary}

Since $H(w)$ and $P(w)$ are conserved in the time evolution, the
Cauchy problem for the averaged NLS equation (\ref{eqn0}) has global
solutions in the energy space $H^1(\mathbb{R}^d)$, if the initial
data $w(x,0)$ gives rise to local solutions in the same energy
space. Therefore, the blowup of solutions of the NLS equation with
$\sigma^2 = 0$ in $d \geq 2$ is arrested by the nonlinearity
management for any $\sigma^2 > 0$.

\section{Averaged Equation versus Full Dynamics}

We have proven that the blowup does not occur in the averaged NLS
equation (\ref{eqn0}) in higher dimensions $d \geq 2$ for $\sigma^2
> 0$. This result raises the question whether the blowup of solutions is arrested in
the full NLS equation (\ref{eqn1}) for any non-zero variance of the
time-periodic nonlinearity coefficient $\Gamma(t)$. We address this
question within the ODE reduction of the time-periodic problem,
which was considered recently with a variational method
\cite{abdul,saito} and a moment method \cite{perez}. In both cases,
the time evolution of the radially symmetric localized solutions of
the full NLS equaiton (\ref{eqn1}) is approximated by a
time-dependent, generalized Ermakov-Pinney \cite{EP} equation:
\begin{eqnarray}
\ddot{R}(t) = \frac{Q_1}{R^3} + \Gamma(t) \frac{Q_2}{R^{d+1}},
\label{eqn5}
\end{eqnarray}
where $R(t) \geq 0$ is an effective width of a localized solution,
while $(Q_1,Q_2)$ are constants found from an initial data, such
that $Q_2 > 0$ (see \cite{perez} for details). We shall consider the
critical case $d = 2$ and rewrite the ODE (\ref{eqn5}) with the
nonlinearity coefficient $\Gamma(t)$ in (\ref{eqn2aa}) in the
explicit form:
\begin{equation}
\label{eqn5a} \ddot{R}(t) = \frac{\alpha + \beta
\gamma(t/\epsilon)}{R^3},
\end{equation}
where $\alpha = Q_1 + \gamma_0 Q_2$, $\beta = Q_2/\epsilon > 0$, and
$\gamma$ is a mean-zero $\epsilon$-periodic function of $t$.
Conditions for blowup and existence of bounded oscillations in
solutions of the ODE (\ref{eqn5a}) were recently reviewed in
\cite{perez} (see also references therein). The sufficient condition
for the blowup (when $R(t) \to 0$ in a finite time $t \to t_0$) is
\begin{equation}
\label{suf-1} \alpha + \beta \max_{0 \leq t \leq \epsilon} (\gamma)
< 0.
\end{equation}
The sufficient condition for the unbounded growth of the solution
width (when $R(t) \to \infty$ as $t \to \infty$) is
\begin{equation}
\label{suf-0} \alpha + \beta \min_{0 \leq t \leq \epsilon}(\gamma) >
0.
\end{equation}
The necessary condition for the bounded oscillations of $R(t) > 0$
for any $t \geq 0$ is
\begin{equation}
\label{suf-2} \alpha < 0, \quad \alpha + \beta \max_{0 \leq t \leq
\epsilon} (\gamma) > 0.
\end{equation}
Numerical simulations in \cite{perez} showed that the condition
(\ref{suf-2}) was also sufficient in the case $\alpha < 0$ for the
blowup arrest for any $t \geq 0$.

It is obvious from the explicit scaling that $\beta \gg |\alpha|$ in
the asymptotic limit $\epsilon \to 0$. Therefore, the second
condition (\ref{suf-2}) is satisfied for $\alpha < 0$ and solutions
of the time-periodic ODE (\ref{eqn5a}) do not collapse, in agreement
with our results derived for the averaged NLS equation (\ref{eqn0}).

Previous work \cite{abdul} (see also the review in \cite{PZ1}) has also
addressed the averaged NLS equation (\ref{eqn0}) in the limit of
{\em weak} nonlinearity management, when $\gamma(\tau)$ is rescaled
as $\gamma = \epsilon \tilde{\gamma}(\tau)$ and the parameter
$\sigma^2$ is small in the limit $\epsilon \to 0$ as $\sigma^2 =
\epsilon^2 \tilde{\sigma}^2$. Solutions of the averaged NLS equation
(\ref{eqn0}) at the leading order $\epsilon = 0$ blow up in a finite
time, but the small $\epsilon^2$-terms formally stabilize the
blow-up for any $\tilde{\sigma}^2 > 0$. Although this approximation
of the critical NLS equation has been considered in many
applications of nonlinear optics (see Sections 4-5 in \cite{FP} and
references therein), it is clearly insufficient for a correct
identification of the domain, where the blowup of solutions occurs.
Indeed, while the averaged NLS equation (\ref{eqn0}) with small
$\sigma^2$ predicts no blowup of solutions, it is clear that the
weak nonlinearity management corresponds to the case $\beta \approx
|\alpha|$ and solutions of the ODE problem (\ref{eqn5a}) (and those
of the full PDE problem (\ref{eqn1})) collapse in the domain
(\ref{suf-1}).

Following the work \cite{saito}, we address the failure of the
averaging procedure for weak nonlinearity management in a simple
time-periodic ODE problem:
\begin{equation}
\label{eqn5b} \ddot{R}(t) = \frac{\alpha + \beta \sin(2 \pi
\tau)}{R^3}, \qquad \tau = \frac{t}{\epsilon},
\end{equation}
where $\alpha < 0$, $\beta > 0$ and $(\alpha,\beta)$ are order of
$O(1)$ in the limit $\epsilon \to 0$. By using the formal asymptotic
multi-scale expansion method (see \cite{PZ1} for details), we
construct an asymptotic solution to the problem (\ref{eqn5b}):
\begin{equation}
\label{series} R = r(t) + \epsilon^2 R_2(\tau,r) + \epsilon^4
R_4(\tau,r) + {\rm O}(\epsilon^6), \qquad \tau = \frac{t}{\epsilon},
\end{equation}
where $R_2$ and $R_4$ are recursively found from the set of linear
inhomogeneous problems,
\begin{eqnarray*}
R_2 & = & - \frac{\beta}{(2\pi)^2 r^3} \sin(2 \pi \tau), \\
R_4 & = & -\frac{3 \alpha \beta}{(2 \pi)^4 r^7} \sin(2 \pi \tau) +
\frac{3 \beta^2}{8 (2 \pi)^4 r^7} \cos(4 \pi \tau).
\end{eqnarray*}
The mean-value term $r(t)$ satisfies an extended dynamical equation
that excludes secular growth of the correction terms of the series
(\ref{series}) in $\tau$:
\begin{equation}
\label{ODE} \ddot{r} = \frac{\alpha}{r^3} + \tilde{\epsilon}^2
\frac{3 \beta^2}{2 r^7} + \tilde{\epsilon}^4 \frac{15 \alpha
\beta^2}{2 r^{11}} + {\rm O}(\tilde{\epsilon}^6), \qquad
\tilde{\epsilon} = \frac{\epsilon}{2 \pi}.
\end{equation}
The averaged ODE problem (\ref{ODE}) is an equation of motion for an
effective particle with a coordinate $r(t)$ in the potential field
with an effective potential energy:
\begin{equation}
\label{potential} U(r) = \frac{\alpha}{2 r^2} + \tilde{\epsilon}^2
\frac{\beta^2}{4 r^6} + \tilde{\epsilon}^4 \frac{3 \alpha \beta^2}{4
r^{10}} + {\rm O}(\tilde{\epsilon}^6).
\end{equation}
When $\alpha < 0$ and $\tilde{\epsilon} = 0$, the particle with
$r(0) > 0$ reaches $r = 0$ at a finite time $t = t_0 < \infty$, that
indicates the blowup of a localized solution. When the next
$\tilde{\epsilon}^2$-term is taken into account (as in the
approximation of weak nonlinearity management \cite{abdul,saito}),
the blow-up is arrested and the mean-value term $r(t)$ oscillates in
an effective minimum of the potential energy $U(r)$, truncated at
$\tilde{\epsilon}^2$-terms. When the next $\tilde{\epsilon}^6$-term
is taken into account (beyond the approximation of weak nonlinearity
management), the potential energy $U(r)$ with $\alpha < 0$ does not
prevent the blowup of the localized solution depending on the initial
data $r(0)$. Existence versus non-existence of blowup depends on the
ratio of parameters $(\alpha,\beta)$ but the difference can only be
detected in the averaging method if convergence of the power series
(\ref{potential}) is established in a closed analytical form.

Similarly, the averaged NLS equation (\ref{eqn0}) can not be used in
the limit of weak nonlinearity management for an accurate prediction
of existence versus non-existence of blowup of solutions. In order
to illustrate this point, we have performed numerical simulations of
the full NLS equation (\ref{eqn1}) in $d = 2$ with
$$
\Gamma(t) = -20.76 + 8 \sin(2\pi t).
$$

\begin{figure}[tbp]
\epsfxsize=10cm 
\centerline{\epsffile{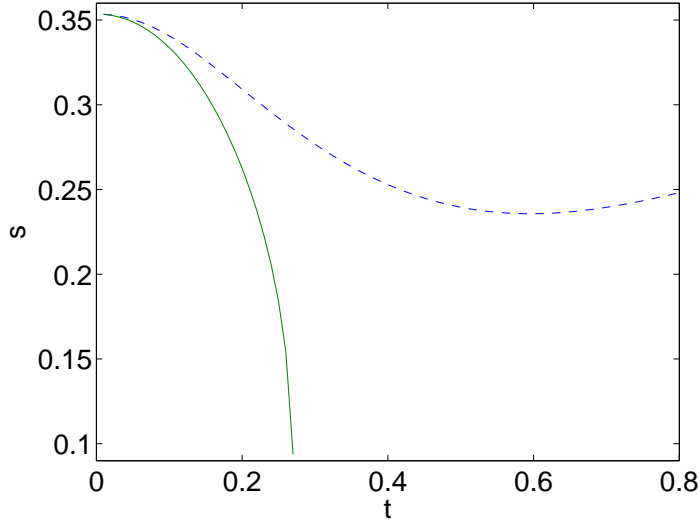}}
\caption{Numerical simulations of the full NLS equation (\ref{eqn1})
(solid curve) and the averaged NLS equation (\ref{eqn0}) with a
fourth order in time scheme, where spacings are $dx=0.075$ and
$dt=10^{-5}$. The half-width $s$ of the wavefunction 
is shown as a function of time $t$. } \label{qfig1}
\end{figure}

We have observed that collapse of localized initial data does occur
(see solid curve on Fig. \ref{qfig1}) by monitoring the half-width 
(of one dimensional slices along $y=0$) of the
wavefunction (for radially symmetric Gaussian initial data),
$$
s = \frac{1}{2} \left( \frac{\int
x^2 |u(x,0,t)|^2 dx
}{\int
|u(x,0,t)|^2 dx} \right)^{1/2},
$$
until it becomes comparable to the lattice grid spacing used (at
that scale collapse is arrested, since the numerical scheme cannot
resolve scales below the grid spacing). On the other hand, numerical
simulations of the averaged NLS equation (\ref{eqn0}) with the same
parameters show that the half-width $s$ never decreased below
$s < 0.23$ (see dashed curve on Fig. \ref{qfig1}) indicating
the absence of collapse in accordance with the rigorous results
presented above. Therefore, the averaged NLS equation (\ref{eqn0})
can only be used for modeling of the blowup arrest in the limit of
strong nonlinearity management of the full NLS equation (\ref{eqn1})
when $\max(\gamma) \gg \epsilon |\gamma_0|$ and $\epsilon$ is small.

\section{Conclusion}

In conclusion, we have studied the global well-posedness of
solutions of the averaged NLS equation that describe strong
nonlinearity management of the time-periodic NLS equation. We have
showed with formal scaling arguments and rigorous analysis that the
blowup of solutions in higher dimensions is arrested within the
averaged NLS equation. We have also discussed the non-applicability of
the averaged NLS equation to the weak nonlinearity management, where
the blowup of solutions can occur beyond the weak management limit.

It is an open problem to study well-posedness of the full
time-periodic NLS equation, depending on parameters of the
nonlinearity management and profile of initial data. Rigorous
results on the latter problem are only available within the ODE
approximation (\ref{eqn5}), when the PDE model is reduced to a
dynamical system with one degree of freedom. It would be
particularly interesting to study mathematically and to examine
numerically whether the theoretical prediction from the method of
moments provides an optimal bound for the full PDE model with arbitrary
initial data.

\vspace{5mm}

{\bf Acknowledgements}. 
The first author is supported in part by NSF-DMS 0204585,
an NSF CAREER award and the Eppley Foundation for Research. The
second  author supported in part by  NSF-DMS 0300511. The third
author is supported by NSERC and PREA. The authors are thankful to
Prof. D.E. Frantzeskakis for numerous enlightening discussions on
the subject.

\end{document}